\documentclass[a4paper, 12pt, amsfonts, amssymb, amsmath, reprint, physics, showkeys, nofootinbib, twoside]{revtex4-2}
\usepackage[utf8]{inputenc}
\usepackage[T1]{fontenc}
\usepackage[english]{babel}
\usepackage[colorinlistoftodos, color=green!40, prependcaption]{todonotes}
\usepackage{amsthm}
\usepackage{mathtools}
\usepackage{xcolor}
\usepackage{graphicx}
\usepackage[left=23mm,right=13mm,top=35mm,columnsep=15pt]{geometry} 
\usepackage{adjustbox}
\usepackage{placeins}
\usepackage{csquotes}
\usepackage{float}
\usepackage{enumitem}
\usepackage[pdftex, pdftitle={Article}, pdfauthor={Author}, colorlinks=true]{hyperref} 
\usepackage{cleveref}

\hypersetup{colorlinks,linkcolor=black,citecolor=black,urlcolor=black} 
\usepackage{natbib}
\begin{document}
\title{Large bubble drives melting in circular DNA}

\author{Souradeep Sengupta, Somendra M. Bhattacharjee, Garima Mishra}
    \affiliation{Department of Physics, Ashoka University, Sonipat, Haryana - 131029, India}

\date{\today} 

\begin{abstract}
We investigate the melting transition of non-supercoiled circular DNA of different lengths, employing Brownian dynamics simulation. In the absence of supercoiling, we find that melting of circular DNA is driven by a large bubble, which agrees with the previous predictions of circular DNA melting in the presence of supercoiling. By analyzing sector-wise changes in average base-pair distance, our study reveals that the melting behavior of circular DNA closely resembles that of linear DNA. Additionally, we find a marked difference in the thermal stability of circular DNA over linear DNA at very short length scales, an effect that diminishes as the length of circular DNA increases. The stability of smaller circular DNA is linked to the occurrence of transient small bubbles, characterized by a lower probability of growth.

\end{abstract}


\maketitle
\section{Introduction}
\label{sec:intro}

DNA serves as the repository of genetic information, 
encompassing the instructions essential for the development, 
survival, and reproduction of all living organisms, 
ranging from the simplest prokaryotes, such as bacteria \cite{roth_evidence_1967, bhagavan_chapter_2002} and viruses \cite{tisza_discovery_2020, vinograd_physical_1966}, 
to the more intricate eukaryotes, including humans \cite{alberts_molecular_2002}. Prokaryotic DNA takes the form of a continuous loop without loose ends, characterized by its circular structure. On the other hand, eukaryotes typically exhibit linear DNA with distinct ends. Additionally, recent research findings indicate that extrachromosomal circular DNAs (eccDNAs) are present independently of linear chromosomal DNA in eukaryotes as well \cite{wahl_importance_1989, yan_current_2020}. The stability of DNA depends on the a number of physical and chemical parameters, such as the pH value of the environment \cite{ageno_alkaline_1969}, salt concentration, temperature \cite{wartell_thermal_1985} as well as mechanical forces \cite{bhattacharjee_unzipping_2000, orlandini_mechanical_2001}. The thermal denaturation transition, i.e. melting \cite{theodorakopoulos_statistical_2020} of DNA, in particular, has been studied for the last 50 years, both in its biological specificity, as well as a statistical and polymer physics problem, due to the greater accessibility of heating experiments \textit{in vitro} and the importance of DNA melting to the polymerase chain reaction \cite{mullis_21_1987}. The melting of linear DNA has been understood theoretically, largely as a sharp first-order phase transition \cite{kafri_why_2000}. The linear DNA  melting is driven by the replication fork or the Y-fork, since the base-pairs at the ends are entropically favoured to open up.  However, the thermal melting of circular DNA is comparatively less understood, although the importance of circular DNA in human pathology \cite{barreto_small_2014} is becoming increasingly clear. 

In the absence of free ends in circular DNA, the melting behavior is controlled by the presence of bubbles. The majority of theoretical studies have investigated circular DNA within the framework of topological constraints, where the twisting and bending dynamics of DNA are inherently interconnected \cite{benham_energetics_1992, jeon_supercoiling_2010, fosado_dynamical_2017, skoruppa_bend-induced_2018, fosado_nonequilibrium_2021}. This is because the Calugareanu-Fuller-White theorem \cite{fuller_decomposition_1978} places a strict constraint on the linking number difference of any circular DNA system. Any flexibility gained by one part of the system by opening up a few base-pairs will have to be compensated by the twist and writhe of other parts of the molecule \cite{morozov_helixcoil_2005}, where the base-pairs are hard to open up. These constraints, relaxed in linear DNA, result in a distinct melting behavior -- the transition in topologically constrained DNA, such as closed circular viral DNA, is far less abrupt compared to that in linear DNA \cite{vinograd_early_1968}. Studies have demonstrated that the competition between melting and local supercoiling induces phase coexistence of denatured and intact phases at the single-molecule level, contributing to the broadening of the transition in circular DNA \cite{benham_torsional_1979, fosado_dynamical_2017}. 

The bubble-mediated melting theory shows that the nature of DNA melting is reliant on the non-extensive logarithmic contribution to the entropy, which is of the form $S=Ns_0 - c \ln{N}$ for a bubble of length $N$ with $s_0$ as the bulk entropy per unit length and $c$ is the reunion exponent for bubble formation \cite{poland_occurrence_1966, poland_theory_1970, fisher_effect_1966, fisher_walks_1984}. In this scheme, a circular DNA under overtwist and supercoiling has been shown to undergo a continuous melting transition with the emergence of a macroscopic loop at $T_c$ \cite{bar_denaturation_2012, kabakcioglu_macroscopic_2012}.
The bubble length  grows monotonically with temperature $T\to T_c^{-}$
until a maximum of half the chain is denatured, as the topological constraints prevent complete melting of the DNA molecule, unlike the linear case. Experiments conducted on supercoiled molecules also indicate that only one denatured bubble occurs per molecule \cite{jeon_supercoiling_2010}. Moreover, no stable bubbles could be detected when supercoiling was absent due to their short lifetime \cite{jeon_how_2008, jeon_supercoiling_2010, sicard_dynamical_2020}, although these measurements were made away from melting due to experimental aims and constraints. In the absence of supercoiling, one would also expect complete denaturation of the molecule, similar to the prediction of a diverging bubble size at criticality in the original Poland-Scheraga (PS) model. 

It is important to acknowledge the existence of circular DNA structures without helical intertwining or topological linkages between strands, such as the singly-nicked polyoma form II DNA \cite{vinograd_physical_1966, vinograd_early_1968} and relaxed plasmid DNA \cite{jeon_supercoiling_2010}, and multiple other studies have both proposed and found evidence for topologically unconstrained closed circular DNA \cite{rodley_possible_1976, wu_novel_1996, biegeleisen_topologically_2002}. These structures provide an opportunity to unravel the effects of two separate kinds of constraints that could be applied to a closed circular DNA: (a) the helical winding of the two strands and the consequent torsional stress and supercoiling, and (b) the topological constraint of closure of the two covalent bonded strands individually while in the bound state. In the absence of a constraint of the first kind, which may be relaxed either via nicking or topoisomerase, the thermal behaviour and sedimentation analysis of circular DNA was at first seen to be very similar to linear DNA \cite{bates_dna_2005, vinograd_early_1968}.
One would generally expect that singly-nicked circular DNA 
will undergo melting due to the fraying of its ends 
originating from the nicked site and results in melting similar to that observed in linear DNA.
However, the specific impact of dual covalent bond closure alone in relaxed circular DNA melting, as compared to linear DNA, remains open for investigation. 

To address this, we use coarse-grained Brownian dynamics simulations to study the melting transition in non-supercoiled circular and linear DNA. Section \ref{sec:methods} describes our model and the details of our simulations. Section \ref{sec:results} describes the melting mechanism employed in circular DNA, and the transient involvement of bubbles in the melting process. We summarize our findings in Section \ref{sec:conclusions}. Some of the details are provided in the Supplementary Information in the form of figures. 

\section{Model and Method}
\label{sec:methods}

We utilize a simplified, low-resolution model 
for homopolymeric DNA composed of two complementary strands 
\cite{mishra_role_2011}. The energy function governing the 
interactions of $N$ base-pair DNA in this model is 
$E_{\mathrm{DNA}} = E_{\mathrm{strand-I}} + E_{\mathrm{strand-II}} + 
E_{\mathrm{strand-I, strand-II}}$. Here, $E_{\mathrm{strand-I}}$ and $E_{\mathrm{strand-II}}$ represent the potential functions characterizing each individual DNA strand, given by
\begin{eqnarray}
 \left. \begin{array}{ll}
 E_{\mathrm{strand-I}}\\ 
 E_{\mathrm{strand-II}}
 \end{array}\right\}& = &  
 \sum_{i=1}^{N-1}K(d_{i,i+1}-d_0)^2 + k_{\mathrm{topo}}(d_{1,N}\nonumber\\ 
 -d_0)^2 + 
& & 4 \sum_{\rm N-nat}\bigg
(\frac{\sigma_{i,j}}{d_{i,j}}\bigg)^{12}
\label{eq:1}
\end{eqnarray}
The distance between beads
$d_{i,j}$ is defined as $|\mathbf{r}_i-\mathbf{r}_j|$, where $\mathbf{r}_i$ and
$\mathbf{r}_j$ are the position vectors of beads $i$ and $j$, respectively
with $i,j\in [1,N]$.  We use dimensionless distances with
$\sigma_{i,j}=1$ and $d_0=1.12$. The energy parameters in the
Hamiltonian are in units of $k_B T$ where $k_B$ is the Boltzmann
constant (set to $1$) and $T$ is the temperature measured in 
reduced units. 
The harmonic spring with dimensionless spring constant $K=100$ 
connecting adjacent beads along the strand, is given by the first 
term on the rhs of Eq. \ref{eq:1}. The second term in Eq. \ref{eq:1} 
with spring constant $k_{\mathrm{topo}}$ is responsible for connecting the 
first and last beads on each strand. The third term introduces a repulsive potential that prevents the overlap of non-native pairs of monomers in strand-I and strand-II \cite{mishra_dynamical_2013, allen_computer_2017, frenkel_understanding_2023}. The two complementary strands of DNA interact via the following potential
\begin{eqnarray}
 E_{\mathrm{strand-I,strand-II}} & = &  \sum_{\rm N. C.}4{\epsilon_{p}}\bigg[\bigg(\frac{\sigma_{i,j}}{d_{i,j}}\bigg)^{12} -
\bigg(\frac{\sigma_{i,j}}{d_{i,j}}\bigg)^{6}\bigg] \nonumber \\
& & + \sum_{\rm N-nat}4\bigg(\frac{\sigma_{i,j}}{d_{i,j}}\bigg)^{12}.\label{eq:2}
\end{eqnarray}
The base pairing between strand-I and strand-II is considered using the first
term on the rhs of Eq. \ref{eq:2} with $\epsilon_p=1$. The native base-pair contacts (same
$i$ of both the chains \cite{mishra_dynamical_2013, mishra_role_2011}) give rise to the ladder like structure of DNA.
The second repulsive term of the potential energy in Eq. \eqref{eq:2}
prevents overlapping of non-native pairs of monomers of 
strand-I and strand-II
\cite{mishra_role_2011, mishra_dynamical_2013, allen_computer_2017, frenkel_understanding_2023}. 

The dynamics of this system is governed by the Brownian equations, given by
\begin{equation}
\frac{d\mathbf{r_i}}{dt} = \frac{1}{\zeta}(\mathbf{F}_c + \Gamma)
\label{Brow}
\end{equation}
Here, $\mathbf{F}_c=-\mathbf{\nabla} E_{\mathrm{DNA}}$ is the conservative force and $E_{\mathrm{DNA}}$ is the sum of $E_{\mathrm{strand-I}}$, $E_{\mathrm{strand-II}}$ and $E_{\mathrm{strand-I,strand-II}}$. $\zeta$ is the friction coefficient, here set to 50, and $\Gamma$ is  the random force, a white noise with zero mean and correlation
$\langle\Gamma_i(t)\Gamma_j(t')\rangle=2\zeta k_B T\delta_{i,j}\delta(t-t')$. These equations of motion are integrated via the Euler method with time step $\delta t = 0.0015$ for $10^8$ iterations at every temperature. To investigate how topology affects the DNA melting process, chains were subjected to topological constraints by setting $k_{\mathrm{topo}}=K=100$, effectively creating circular DNA. The melting of linear DNA, which lacks topological constraints ($k_{\mathrm{topo}}=0$) has been studied extensively \cite{azbel_phase_1979, theodorakopoulos_order_2000, zeng_bubble_2004, ares_bubble_2005, theodorakopoulos_dna_2008, rieloff_structural_2017, majumdar_softening_2020} and provides a platform for investigating the influence of topology on the melting of DNA.

\begin{figure}[t]
\includegraphics[width=0.48\textwidth] {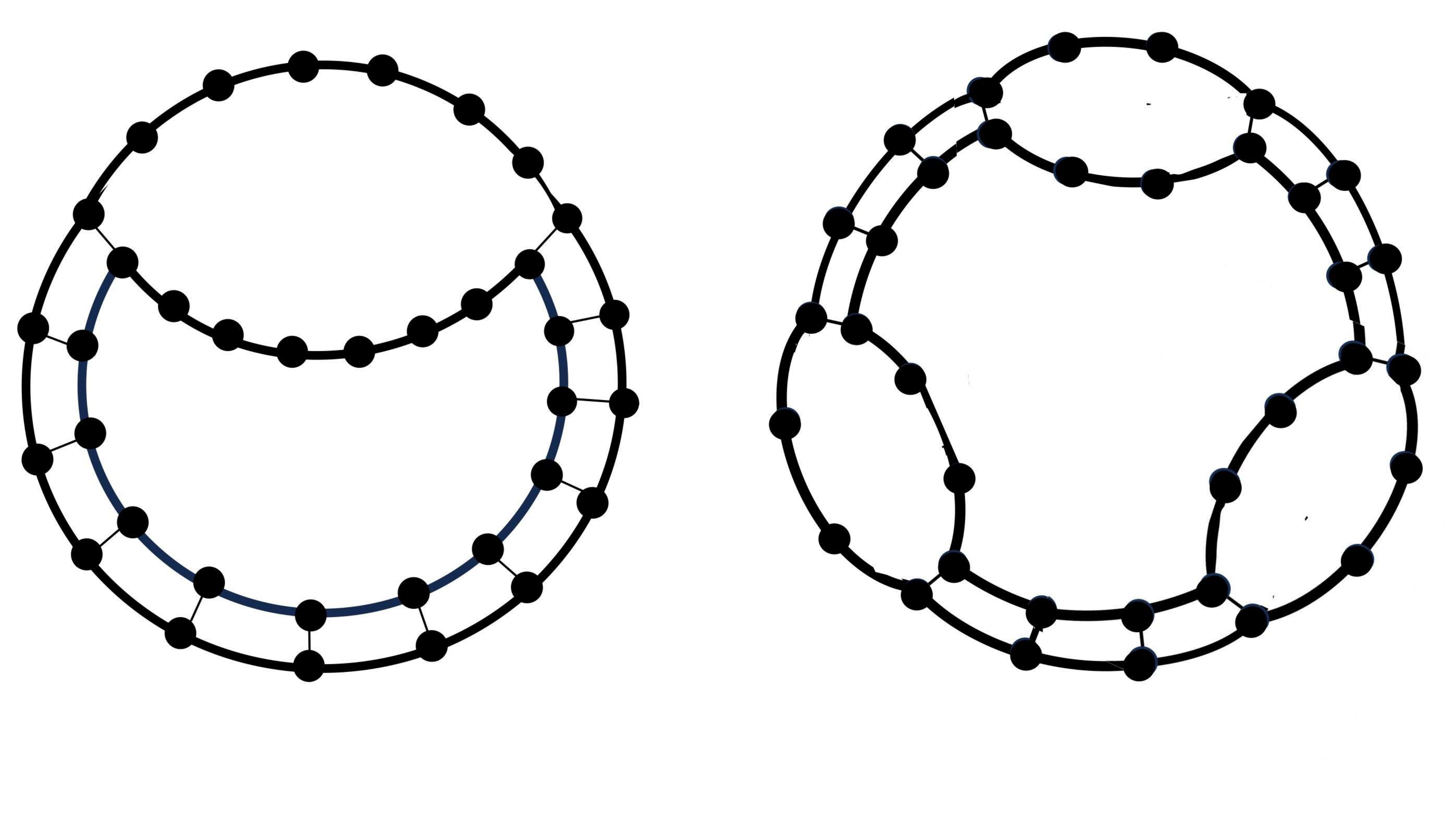}
\caption{Proposed schematic diagrams for melting in non-supercoiled circular DNA: (a) Type-I, involving a single large loop, and (b) Type-II, incorporating multiple small loops. Here, thick lines denote covalent intra-chain bonds, while thin lines denote inter-chain native base-pairing hydrogen bonds.}
\label{fig:schematic}
\end{figure}

In the absence of a free end, circular DNA can undergo melting through the formation of bubbles. To obtain a microscopic view of the bubbles involved in the melting of circular DNA devoid of supercoiling, we partitioned the entire chain into three equal sections and monitored sector-wise change in average base-pair distance, denoted as $r_1$, $r_2$ and $r_3$, relative to the initial bound  state. 
The sector wise measurements enable the detailed characterization of large or small bubbles within the system.
If one or two segments play a substantial role, with $r_1$ or $r_2$ assuming a significant magnitude, while $r_3$ takes on a comparatively smaller value, an anticipated structure involves
a large bubble, resembling the configuration in Fig. \ref{fig:schematic}a. Conversely, if each segment contributes roughly equally i.e. $r_1 \approx r_2 \approx r_3$, the resulting structure is multiple small bubbles distributed uniformly along the chain in Fig \ref{fig:schematic}b. Hence, the melting of circular DNA can be characterized by two types of possible pathways : (i) Type-I, marked by the emergence of a large bubble, and (ii) Type-II, where small bubbles uniformly form across the entire chain. As temperature increases, these bubbles grow, which eventually leads to the melting of the whole circular DNA. It would be interesting to investigate whether 
the melting of circular DNA (without supercoiling) follows 
the Type-I pathway (distinguished by a large bubble) as proposed in earlier studies for supercoiled DNA \cite{fosado_dynamical_2017, jeon_supercoiling_2010, kabakcioglu_macroscopic_2012} or if it employs the Type-II pathway, especially in the presence of small bubbles.

\section{Results}
\label{sec:results}

\subsection{Melting and Sectorwise Basepair Opening}
\label{sec:results-sector}

In the context of DNA melting, the fraction of fully bound 
base pairs $N_b$ serve as an order parameter. In the 
fully bound state, where all base pairs remain intact, 
$N_b$ takes on a value $\approx 1$.  As the temperature 
increases, the fraction of fully bound base pairs decreases, indicating a transition toward 
a more disordered (unbound) state. A base-pair is classified as bound when its native base-pair-distance is below the standard cutoff distance of 1.5. A completely unbound state is 
characterized by $N_b=0$. The widely
used definition for melting point $T_m$ is a 
temperature at which half of the base-pairs of the DNA 
chain are open. The peak position in the specific heat vs temperature curve also gives the melting temperature ($T_m$). Here, we examined the melting of DNA chains of different lengths (18bp, 36bp, 72bp, and 144bp), both in the presence and absence of the constraint $k_{\mathrm{topo}}$. 
We observe that the fraction of fully bound base pairs goes from $\approx$1 to 0 (see Figure S1 in Supplementary Information). We note that the melting temperature ($T_m$) of circular 
DNA is noticeably higher than that of linear DNA when considering a length of $18$bp and $36$bp. Nevertheless, as the chain size increases ($72$bp, $144$bp), this distinction in $T_m$ for circular and linear DNA appears to diminish (see Figures S1 and S2 in Supplementary Information). Although both ends of DNA strands are covalently linked in our model, our findings are nevertheless in accordance with experimental observations that circular polyoma form II viral DNA (singly nicked) shares the same melting temperature as linear T7 DNA \cite{vinograd_early_1968}, since both systems are torsionally relaxed and therefore devoid of supercoiling. 

\begin{figure}[t]
    \centering
   \includegraphics[width=0.48\textwidth]{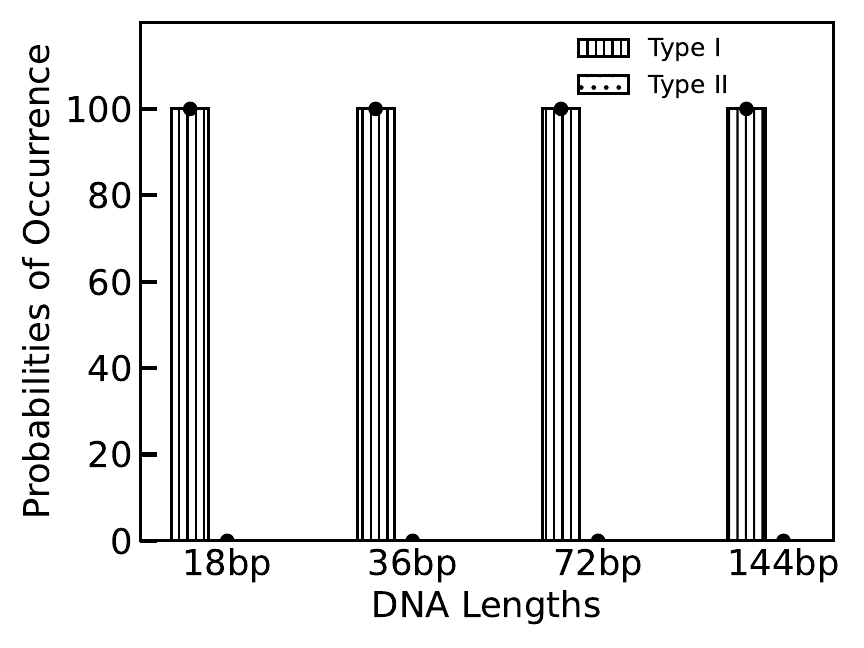}
   \caption{ Histogram illustrating the melting behaviour of circular DNA categorized through Type-I and Type-II scheme. Each circular DNA molecule is divided into three equal sectors, and the average contribution of each sector to the total change in base-pair distance is calculated in the transition region.   
   For a Type-I transition, a single sector    
  alone contribute more than $40\%$ of the total (signifying a big bubble), while for Type-II transition, each sector is expected to contribute more than $30\%$
 (signifying many distributed small bubbles). The melting of non-supercoiled circular DNA is predominantly occurs through Type-I transition.}
    \label{fig:circhist}
\end{figure}

\begin{figure*}
   \includegraphics[width=0.95\textwidth]{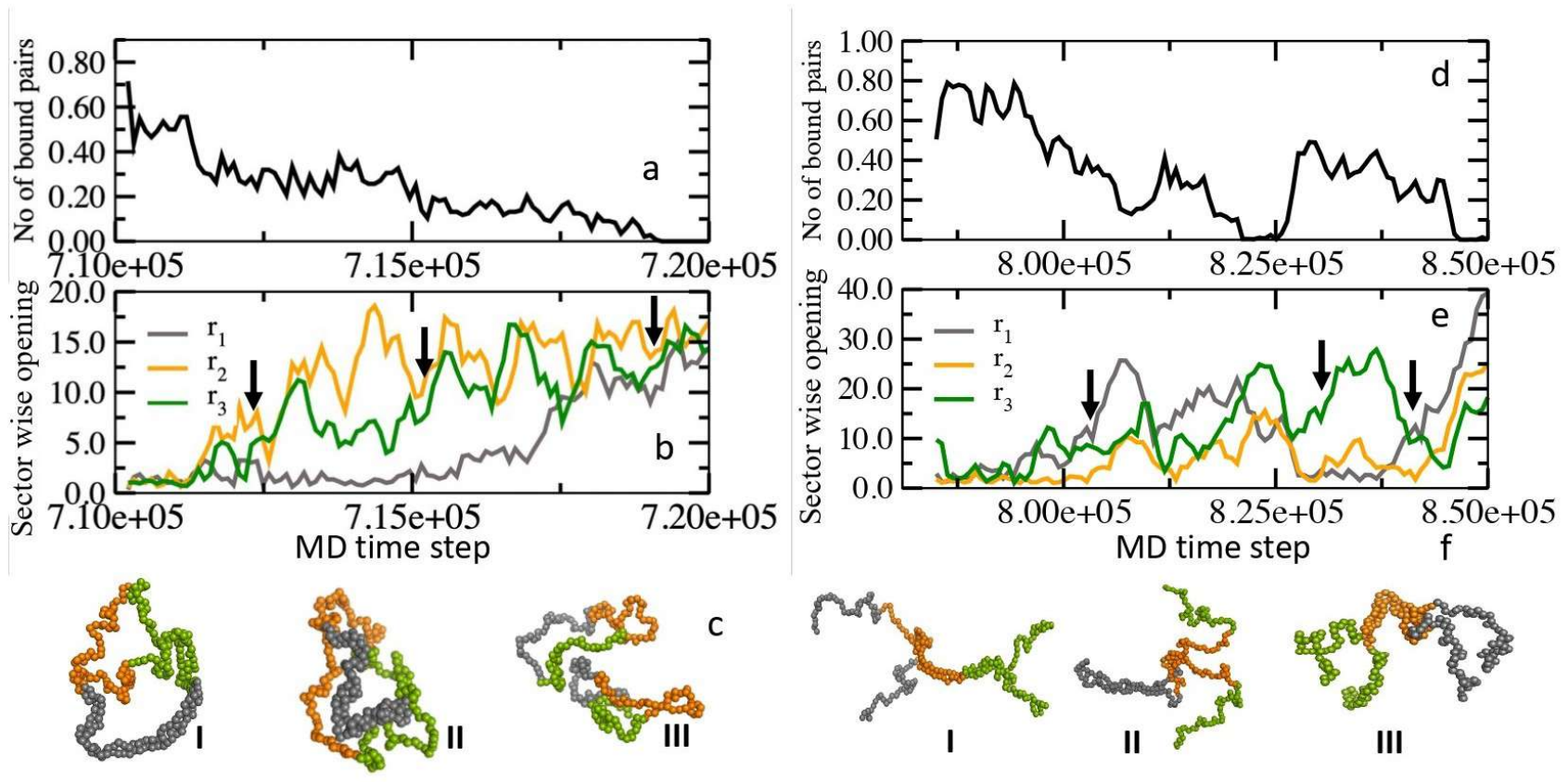}
   \caption{Comparison of opening mechanism between circular and linear DNA. (a, d) Illustrates the time evolution of the number of bound base-pairs. (b, e) Depicts the sector-wise change in base-pair distance. (c, f) Snapshots of the opening process at different time steps shown by arrow.}
    \label{fig:mechanism}
\end{figure*}

In order to probe the melting mechanism, we characterize the Type-I and Type-II transitions quantitatively.
The Type-I transitions require that, at 
least one sector should contribute, on average, approximately $> 40\%$ of the total change in average base-pair-distance $R (=r_1+r_2+r_3$), while the combined contribution of the other two sectors should be approximately $\le 60\%$ of the R in the transition region. The transition region is defined as the region where the number of intact base-pairs decreases from $80\%$ to $0\%$. In a Type-II transition, each sector must contribute over $\sim 30\%$ of the R for the chain in the transition region. We conduct 100 independent simulations each, for both circular and linear DNA of different lengths, and classify each trajectory using the same criteria for all lengths. The results are shown in Fig. \ref{fig:circhist}.
Our findings suggest that the melting behaviour of circular DNA ($18$bp, $36$bp, $72$bp and $144$bp) is predominantly characterized by Type-I transitions, marked by the formation of a substantially large bubble. We also investigated the melting behavior of linear DNA by employing sector-wise change in average base-pair distance measures. We observe that the melting process of linear DNA is also governed by Type-I transitions (see Figure S3 in Supplementary Information).

To explore the microscopic picture in Type-I pathway, we examine representative trajectories for circular and linear DNA ($144$bp), depicted in Fig. \ref{fig:mechanism}.  The three sectors of circular/linear DNA are represented using three distinct colors: grey, orange, and green.
At the onset of the transition, most base-pairs remain intact, with the total number of bound base pair fraction close to 1 (Fig \ref{fig:mechanism}a). As expected, the corresponding sectorwise changes in average base-pair-distance $r_1$, $r_2$, and $r_3$ are approximately 0 (Fig \ref{fig:mechanism}b). The resulting structure corresponds to bound circular DNA.
As time progresses, the fraction of intact base pairs starts decreasing
(Fig \ref{fig:mechanism}a). The $r_2$ begins to exhibit an increase in magnitude, followed by a rise in $r_3$, while $r_1$ remains near zero (Fig \ref{fig:mechanism}b).
A  bubble emerges in the orange sector and grows substantially (snapshot I in (Fig \ref{fig:mechanism}c), characterized by a large $r_2$,  while a smaller bubble forms in the green sector, exhibiting a slightly reduced $r_3$ with unchanged $r_1 \approx 0$ i.e. intact grey sector. As time progresses, the bubble in green sector also grows along with bubble in orange sector(snapshot II in (Fig \ref{fig:mechanism}c). At later times, the number of bound base pairs approaches zero, $r_2$ and $r_3$ approach their maximum values, finally followed by an increase in $r_1$ as well (Fig \ref{fig:mechanism}a-b). This continues until the whole DNA gets separated (snapshot III in Fig \ref{fig:mechanism}c).
In linear DNA, at the onset of the transition region, all base-pairs remain intact (Fig \ref{fig:mechanism}d), and the breakage of base pairs initiates from the free ends \ref{fig:mechanism}e, 
\ref{fig:mechanism}f), due to high end-entropy. As time progresses, a large Y-fork originates in the grey and orange sectors (see snapshot-I Fig \ref{fig:mechanism}f) with some broken base-pairs in the green sector as well, representing a smaller Y-fork at the opposite end of the linear DNA. The change in average base-pair distance and fraction of bound base-pairs plots (Fig. \ref{fig:mechanism}d, e) also indicate the closure of one Y-fork at later times and large Y-fork at other end (snapshot-II in Fig. \ref{fig:mechanism}f), as evidenced by smaller values of $r_2$ and $r_1$. Therefore, the melting process of linear DNA occurs through the Y-forks, as previously investigated \cite{majumdar_softening_2020, theodorakopoulos_statistical_2020, kafri_why_2000}.

\begin{figure}[t]
 \centering
\includegraphics[width=0.48\textwidth]{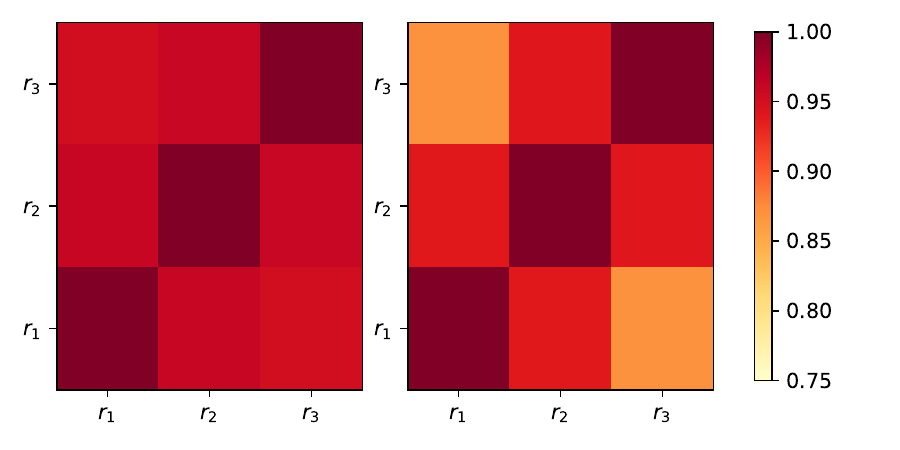}
   \caption{Time-averaged and ensemble-averaged (over all independent simulations) correlation map for the sectorwise change in average base-pair distance $r_1, r_2, r_3$ for circular (left) and linear (right) DNA of length 144bp. A large bubble can emerge anywhere along the circular DNA, resulting in a more uniform correlation map. In linear DNA, the lack of correlation between the two ends indicates that melting primarily occurs due to the gradual opening of a single Y-fork from either end.}
    \label{fig:corr}
\end{figure}

We also sought correlations between $r_1$, $r_2$ and $r_3$ to elucidate the statistical behavior of DNA melting for both circular and linear DNA. The correlation between different sectors shows uniformity (Fig. \ref{fig:corr}a) for circular DNA.  The melting process in circular DNA is characterized by the emergence of large bubbles in two adjoining segments. Since these large bubbles can emerge between any two segments, the time- and ensemble-averaged correlation map will be uniform overall. In the case of linear DNA, the correlation map indicates that the two ends are less correlated compared to the adjacent sectors (see Fig. \ref{fig:corr}b). A Y-fork structure can form at either end of DNA due to its symmetry, and the formation is equally likely from any side, but not usually simultaneously in both (see Figure S4 of Supplementary Information). Therefore, melting of linear DNA usually takes place by utilizing single large Y-fork.

\subsection{Stable Bubbles and Typical Behaviour}
\label{sec:results-stablebubs}

The fixed linking number constraint inherent in supercoiled circular DNA may lead to the dismissal of the possibility of Type-II melting therein. Our current investigation into circular DNA devoid of supercoiling also does not reveal the presence of Type-II melting. This prompts an inquiry into why Type-II melting, typically associated with multiple small bubbles, is absent in circular DNA devoid of supercoiling.

\begin{figure}[t]
   \includegraphics[width=0.48\textwidth]{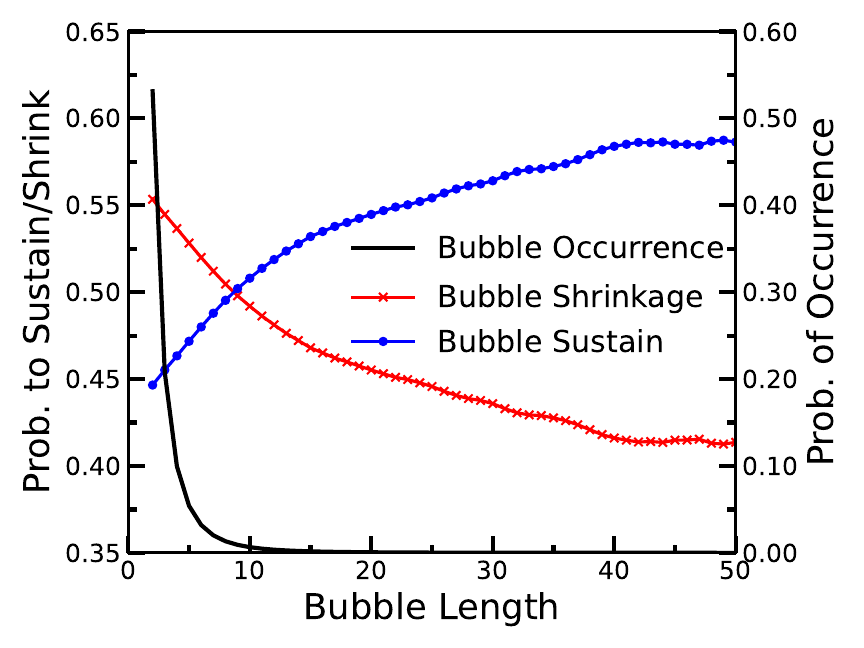}
   \caption{Shrinkage, sustenance (growth/maintenance) and occurrence probabilities for bubbles of different lengths in 144bp circular DNA, averaged over all simulations. 
   At each time-step, if a bubble of length l is observed, it is tracked in the subsequent time-step, and its length is recorded, which over time gives a behavior probability profile for bubbles of different lengths. A crossover at $50\%$ between shrinkage and sustenance probabilities is noted for bubbles of length 10bp, indicating the stable bubble nucleation threshold for this system. Please see Fig. S5 in Supplementary Information.}
    \label{fig:growthprobs144}
\end{figure}

To address this, we extend an algorithm by Hillebrand \textit{et. al.} \cite{hillebrand_distributions_2020} and examine the probability of occurrence of bubbles of different sizes in circular DNA of various lengths ($18$bp, $36$bp, $72$bp, $144$bp). Our findings indicate that smaller bubbles exhibit a higher occurrence probability compared to larger bubbles as shown in by solid black line in Fig. \ref{fig:growthprobs144}. However, it's important to note that static measurements of bubble size provide no insight into the dynamics of bubbles, such as whether they will grow or shrink over time.  For that, we track the dynamical behaviour of bubbles over time. If a bubble of a given size exists at time $t$, we examine what this bubble does at the next recorded time-step $t+\delta t$ (for us, $\delta t=10^3$ timesteps) -- does it grow, does it shrink or does it stay the same length?

To do this, we set up three counters $G(l), M(l), S(l)$ for any bubble of length $l$ (i.e. $l$ consecutive broken bonds) -- these three counters denote the number of growing, maintaining or shrinking events for any such bubble. We track every instance of a bubble of length $l(t)$ appearing at time $t$, and the subsequent behaviour of this bubble at time $t+\delta t$. We consider a bubble at time $t$ to be the same bubble at $t + \delta t$ if they share any bases.
If the length of the bubble decreases, $l(t)>l(t + \delta t)$, we increment $S(l)$ by 1. If the length remains the same, we increment $M(l)$ by 1. If the length increases, 
$l(t) <l(t+\delta t)$, we increment $G(l)$ by 1. Over an entire simulation, a bubble of size $l$ appears $n_l$ times, and $n_{\mathrm{tot}}=\sum_{l} n_l$ gives the total number of occurrences of bubbles of any length throughout the chain. The ratio of $n_l$ to $n_{\mathrm{tot}}$ gives us the probability, or relative likelihood, of the occurrence of a bubble of length $l$ in the simulation, while the ratio of the number of growing ($G(l)$), maintaining ($M(l)$) or shrinking events ($S(l)$) to the total number of occurrences of this bubble ($n_l$) gives us a statistical picture of its typical behaviour, which is then further averaged over all our simulations. 

Our analysis shows that larger bubbles are less likely to occur compared to smaller ones (solid black line in Fig \ref{fig:growthprobs144}),
but once they do, they tend to persist or expand, while smaller bubbles are more likely to shrink 
(blue and red lines in Fig \ref{fig:growthprobs144}). 
By comparing the probabilities of bubble sustenance and shrinkage, we can find the stable bubble nucleation threshold lengths, with crossover around $50\%$ for certain bubble lengths, (see Figure \ref{fig:growthprobs144}). Beyond this length, bubbles are more likely to be stable and long-lived, and therefore contribute to the melting of the whole molecule. We find the threshold size for bubble growth to be $\approx 10$bp. This is comparable to earlier studies \cite{zeng_bubble_2004, ares_bubble_2005, alexandrov_toward_2009, rieloff_structural_2017}, that found nucleation thresholds to be around $\sim 12$--$20$bp for bubbles trapped between bound double-stranded sections in linear DNA (comparable to bubbles forming in closed circular DNA).

The statistical behaviour for various bubble sizes indicates that although multiple small bubbles (approximately $l< 10$) may emerge at different times throughout the entire chain with  a higher likelihood (as would be required for a Type-II transition), the simultaneous presence of a uniform distribution of large bubbles ($l>10$) along the chain is very unlikely, assuming that the behavior of well-separated bubbles can be considered as independent events. Thus, while Type-II melting may be anticipated for smaller-size bubbles, it is not supported by observations, as these bubbles are more likely to shrink. Ultimately, the melting of circular DNA is driven by a large bubble (Type-I), given that the homogeneous distribution of large bubbles all along the chain (Type-II) is less probable.

These results also clarify the effect of system size on the bubble dynamics in circular DNA. In smaller circular DNA molecules ($18$bp and $36$bp), the threshold size for bubble formation is approximately $10$bp, constituting around $40-50\%$ of the total chain length. Bubbles smaller than this threshold size may form, but are likely to quickly shrink. Thus, increasing the temperature until a relatively larger bulge forms is necessary to initiate melting, explaining the requirement for a higher melting temperature ($T_m$) for $18$bp and $36$bp circular DNA, compared to linear DNA of the same size. However, for the larger systems we studied ($72$bp, $144$bp), a bubble of size $10$bp, i.e. $\sim 7-10\%$ of the total length, is stable, can grow, and will lead to melting of the whole chain, similar to linear DNA, where Y-fork openings at the ends have a low nucleation threshold and initiate melting\cite{zeng_bubble_2004, ares_bubble_2005}(see Figure S6 in Supplementary Information for more details). As we increase the system size, the effect of the topological constraint is progressively washed out, and circular DNA starts to behave more like linear DNA. 

\section{Conclusions}
\label{sec:conclusions}

In this paper, we investigated the thermal melting of non-supercoiled circular and linear DNA of four different lengths using Brownian dynamics simulations across a range of temperatures. Our study revealed that covalently closed circular DNA exhibits a higher melting temperature than linear DNA, particularly at smaller length scales. In circular DNA, we identified a critical minimum length threshold for bubbles, which is crucial for their growth or maintenance. Bubbles shorter than this specific threshold tend to shrink rather than maintain stability or expand. This result has significant implications as shorter DNA chains can only accommodate small bubbles, rendering the latter susceptible to shrinkage. As a result, higher temperatures are required to facilitate the  formation of larger bubbles within smaller DNA chains, ultimately leading to the chain's melting. Unlike circular DNA, there is no such restriction on threshold length scale in linear DNA. Our findings also confirmed that the melting of linear DNA occurs through Y-forks originating from either end.  However, there is a weak correlation between the Y-forks at opposite ends of linear DNA, meaning that only one side predominates at any one time. We also ruled out the possibility of melting of non-supercoiled circular DNA via the proliferation of bubbles distributed along the chain. We reaffirmed that the melting of circular DNA predominantly occurs through the formation of one large bubble due to the significantly lower probability of occurrence of multiple threshold-sized bubbles along the chain compared to smaller bubbles.

\section*{Acknowledgements}
GM gratefully acknowledges the financial support from SERB India for a
start-up grant with file Number SRG/2022/001771. SS acknowledges the financial support from Ashoka University and Ashoka's High-Performance-Computing Cluster. 

\bibliography{bibliography}

\pagebreak

\widetext 

\begin{center}
\textbf{\large Supplementary  Information}
\end{center}
\setcounter{equation}{0}
\setcounter{figure}{0}
\setcounter{table}{0}
\setcounter{page}{1}
\setcounter{section}{0}
\renewcommand{\theequation}{S\arabic{equation}}
\renewcommand{\thefigure}{S\arabic{figure}}
\renewcommand{\thesection}{S\arabic{section}}

\begin{center}
{\large{\bf  ``Large bubble drives melting in circular DNA''}}\\
{Souradeep Sengupta}, {Somendra M. Bhattacharjee} and {Garima Mishra}\\
{Department of Physics,  Ashoka University, Sonepat 131029, India}\\

Email: garima.mishra@ashoka.edu.in
\end{center}

\label{app:meltprof}

\begin{figure*}[htb!]
    \includegraphics[width=\textwidth]{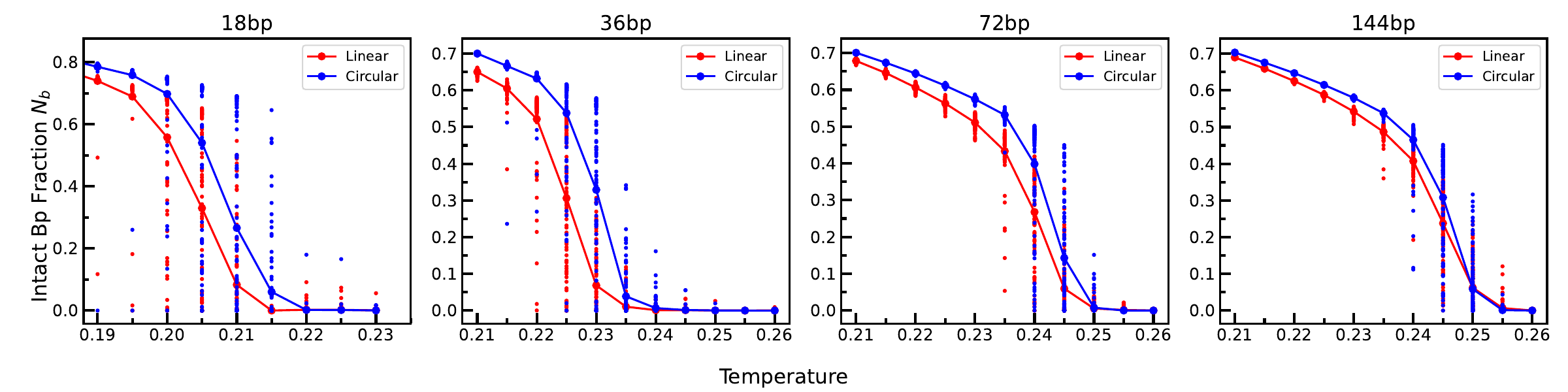}
    \caption{Fraction of intact (fully closed) basepair fraction vs temperature across all lengths. The individual points represent 100 different independent trajectories, while the solid line connects, as a guide to the eye,  the averages of those 100 data-points at every temperature.  While there is a clear distinction between the two for shorter lengths, the melting profiles approach each other with increasing size -- this suggests that the effect of the topological constraint is mediated by scale. These graphs are echoed by the melting profile seen in terms of conformational energy vs temperature, seen in Figure S2.}
    \label{fig:intactbp}
\end{figure*}

\begin{figure*}[!htb]
\includegraphics[width=\linewidth]{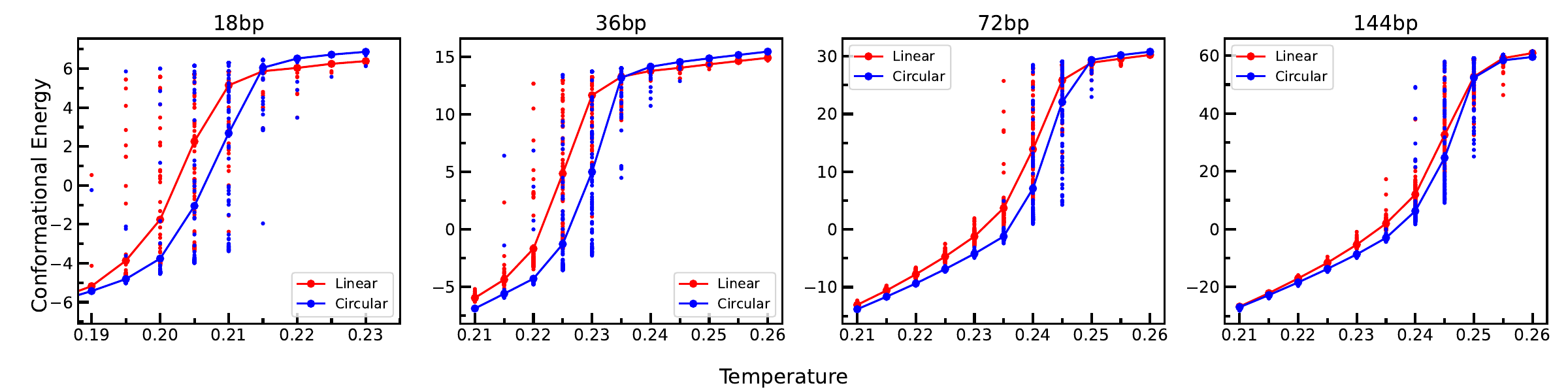} 
\label{fig:EvsT}
\caption{Conformational energy vs temperature across DNA lengths. As in Figure S1, the individual points represent 100 different independent trajectories, while the solid line connects, as a guide to the eye, the averages of those 100 data-points at every temperature.  We see the same behaviour of a distinction between circular and linear DNA becoming less prominent with increasing system size. }
\label{fig:EvsT_profile}
\end{figure*}

\clearpage

\label{app_freehist}

\begin{figure}[htb!]
    \centering
    \includegraphics[width=0.45\textwidth]{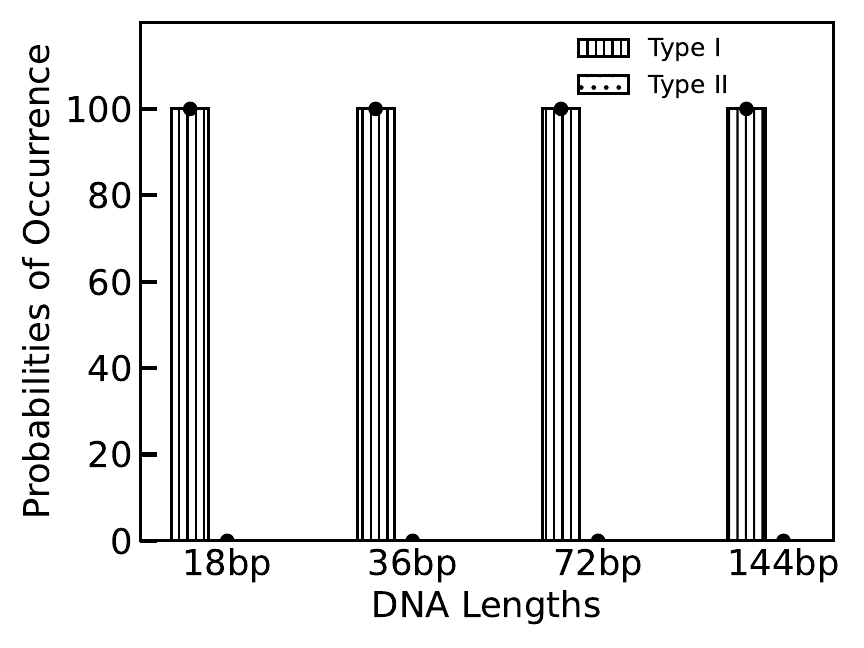}
    \caption{ Histogram illustrating the melting behaviour of linear DNA categorized through type-I and type-II scheme. We divide each DNA into three equal sectors and average the contribution of each sector to the total change in base-pair distance in the transition region. For a type-I scheme, we might expect at least one sector alone to contribute more than $40\%$ of the total (signifying a big bubble), while for type-II, we might expect each sector to contribute more than $30\%$ each (signifying many distributed small bubbles). As we can see, Type-I transitions predominantly account for melting in linear DNA, mediated by large openings (Y-forks) at the two ends.}
    \label{fig:freehist}
\end{figure}

\clearpage

\label{app_corrplot}

\begin{figure*}[htbp!]
    \centering
    \includegraphics[width=\textwidth]{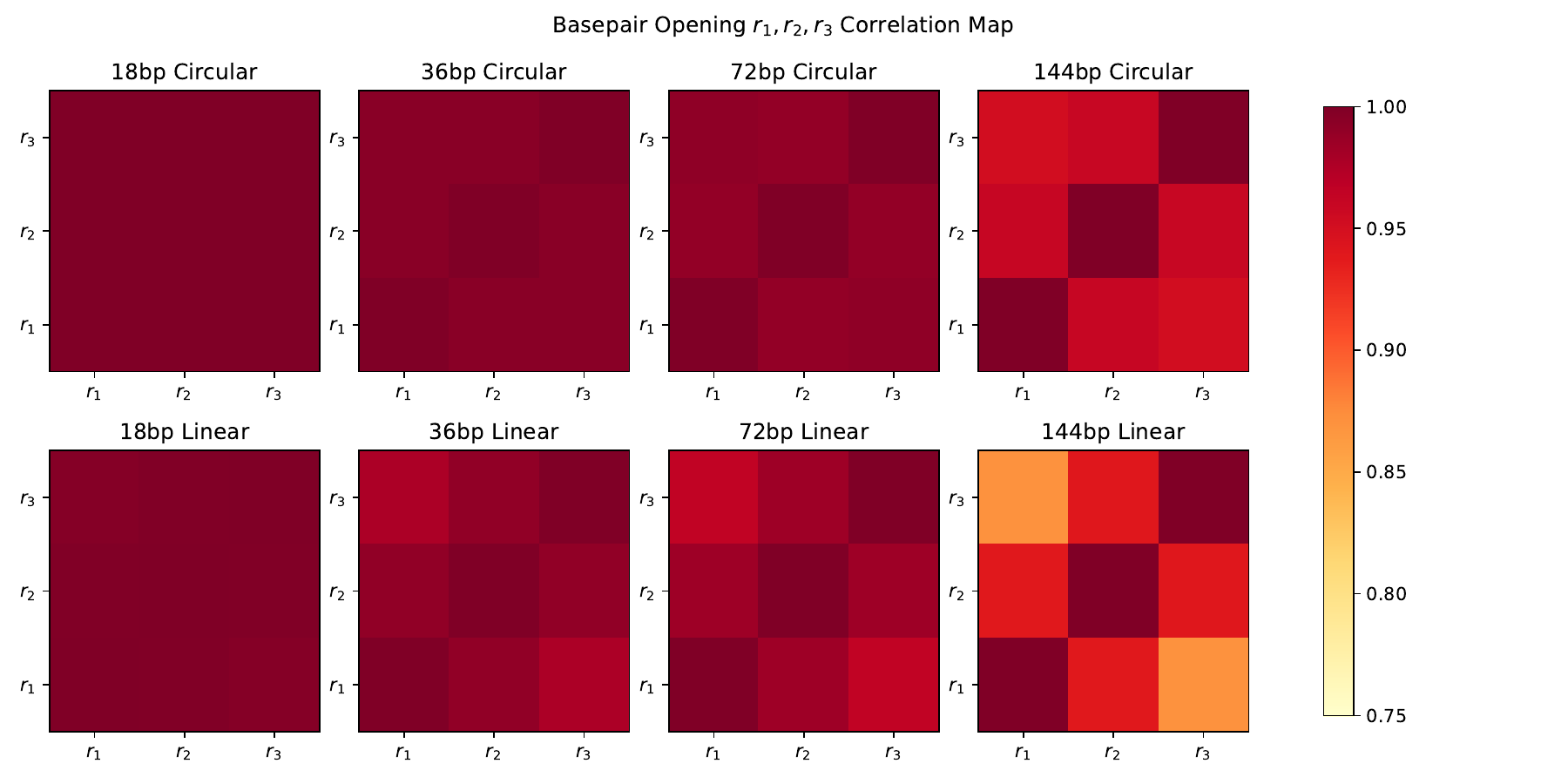}
    \caption{Time-averaged and ensemble-averaged (over all independent simulations) correlation map for the sectorwise change in average base-pair distance $r_1, r_2, r_3$ for circular (top row) and linear (bottom row) DNA of different lengths. Since bubbles can emerge at any point in the circular DNA molecule, the correlation maps on the top are much more homogenous. Whereas for linear DNA, the two ends are relatively uncorrelated and the correlation decays along every row and column starting from the principal diagonal. This indicates that melting in this system is driven primarily by a single Y-fork opening up from either end, rather than by two Y-forks meeting in the middle.}
    \label{fig:rmsd_corrmap}
\end{figure*}

\clearpage

\label{app_growthprobs}

\begin{figure*}[hbtp]
    \centering
    \includegraphics[width=\textwidth]{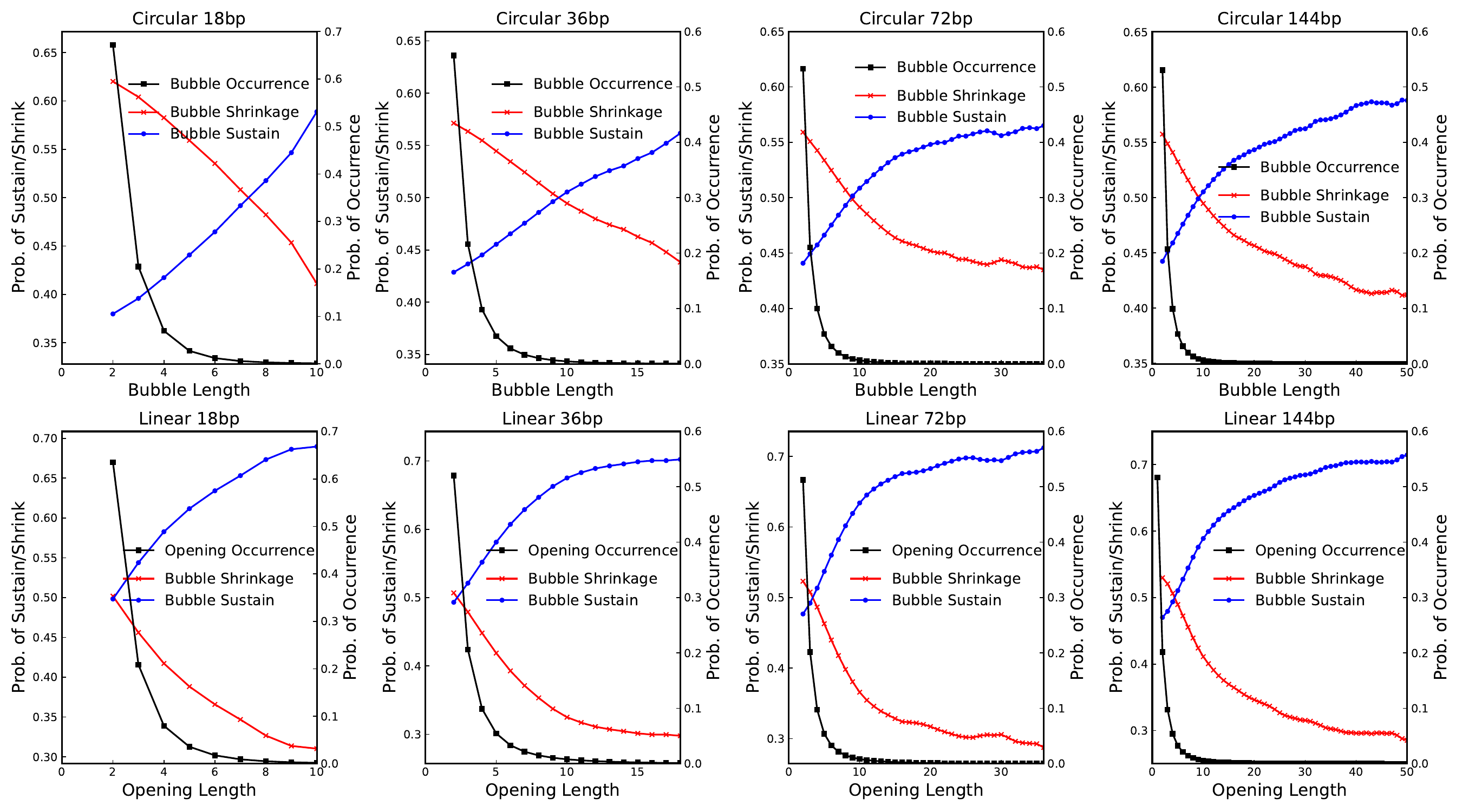}
    \caption{Shrinkage, sustenance (growth/maintenance) and occurrence probabilities for bubbles of different lengths in circular (top row) and linear (bottom row) DNA of all lengths, averaged over all simulations. At any time-step, if we see a bubble of length $l$, we track it at the next time-step and record its length, which over time gives us a behaviour probability profile for bubbles of different lengths. A crossover happens at $50\%$ between the shrinking and sustaining probabilities at a certain bubble length -- this is the stable bubble nucleation threshold. Beyond this length, bubbles are more likely to grow and contribute to chain melting. For circular DNA, this threshold length varies around 8-10bp, while for linear DNA, it varies around 2-5bp, for different DNA sizes.}
    \label{fig:bubgrowth}
\end{figure*}

\begin{figure*}[hbtp]
    \centering
    \includegraphics[width=0.8\textwidth]{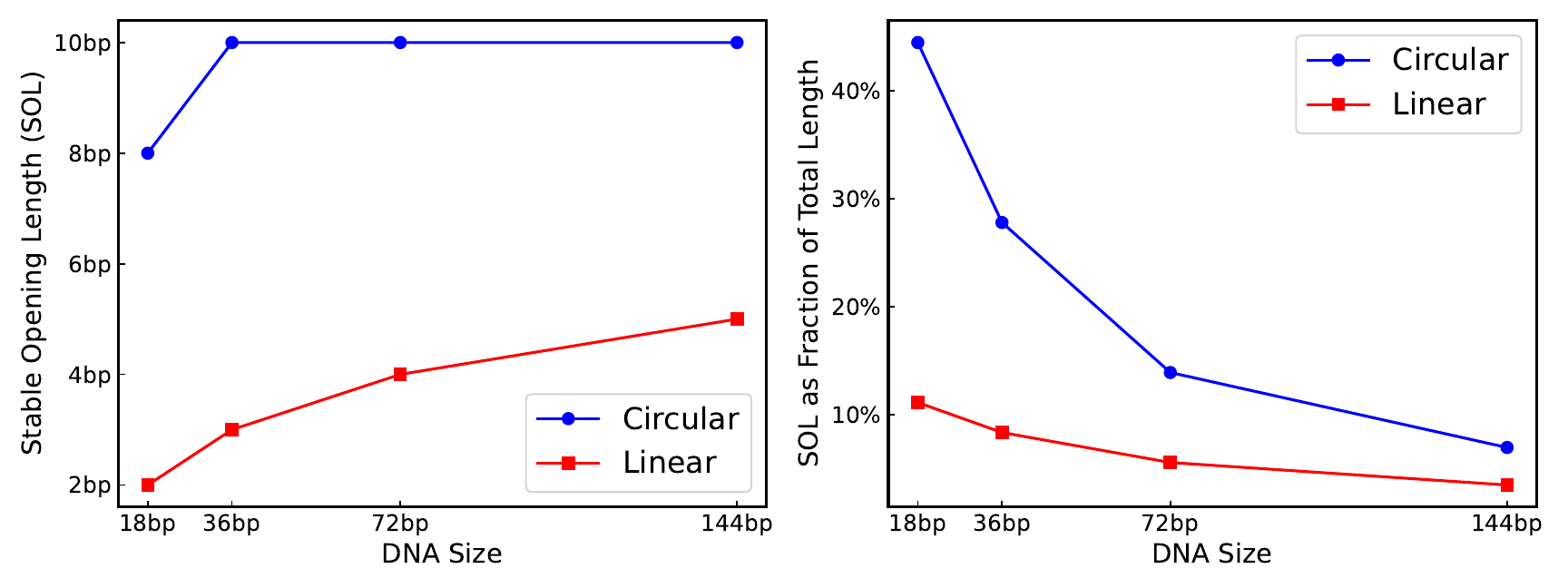}
    \caption{Stable opening nucleation thresholds for circular and linear DNA of different sizes, plotted in absolute terms (left), and  as a percentage of the total DNA length (right). By checking the crossover between sustenance and shrinkage probabilities from FIG. S5, we determined, for each system, the minimum bubble/opening size that is more likely to grow or maintain its size rather than shrink. A smaller length threshold for stable openings indicates that the system is easier to melt, and smaller thermal excitations can stabilise to become self-sustaining. When this minimum opening size is plotted as a fraction of the total length, we find that there is a noticeable difference between the stability of these bubbles/openings between circular and linear DNA for smaller molecules, and this difference reduces with increasing system size, explaining the difference in melting profiles of circular and linear DNA for smaller molecules.}
    \label{fig:nucthresh}
\end{figure*}

\end{document}